\documentclass{egpubl}
\usepackage{eurovis2026}

\preprint

\CGFccby

\usepackage[T1]{fontenc}
\usepackage{dfadobe}  

\usepackage{cite}  
\BibtexOrBiblatex
\electronicVersion
\PrintedOrElectronic
\ifpdf \usepackage[pdftex]{graphicx} \pdfcompresslevel=9
\else \usepackage[dvips]{graphicx} \fi

\usepackage{egweblnk}

\usepackage{blindtext}
\usepackage{xcolor}
\usepackage{amsmath}
\usepackage{amsfonts}
\usepackage{amssymb}
\usepackage{enumitem}

\usepackage{lipsum}
\usepackage{xcolor}
\usepackage{xspace}

\newcommand{\eg}{\emph{e.g.}\xspace}

\newcommand{\myparagraph}[1]{\smallskip\noindent\textbf{#1}\xspace}

\definecolor{myhighlight}{HTML}{FF7F00}

\newcommand{\systemname}{\textsc{UrbanClipAtlas}\xspace}

\title[\systemname]%
      {\systemname: A Visual Analytics Framework for Event and Scene Retrieval in Urban Videos}

\author[J. Perca et al.]
{\parbox{\textwidth}{\centering Joel Perca$^{1}$\orcid{0009-0000-4917-4747}, Luis Sante$^{1}$\orcid{0009-0009-6547-7313}, Juanpablo Heredia$^{1}$\orcid{0000-0002-6126-4881}, Joao Rulff $^{1,2}$\orcid{0000-0003-3341-7059}, Claudio Silva$^{2}$\orcid{0000-0003-2452-2295}  and Jorge Poco$^{1}$\orcid{0000-0001-9096-6287} 
        }
        \\
{\parbox{\textwidth}{\centering $^1$Fundação Getulio Vargas, Brazil\\
         $^2$New York University, USA
       }
}
}

\begin{document}

\teaser{
 \centering
  \vspace{-0.9cm}
  \includegraphics[width=0.92\linewidth]{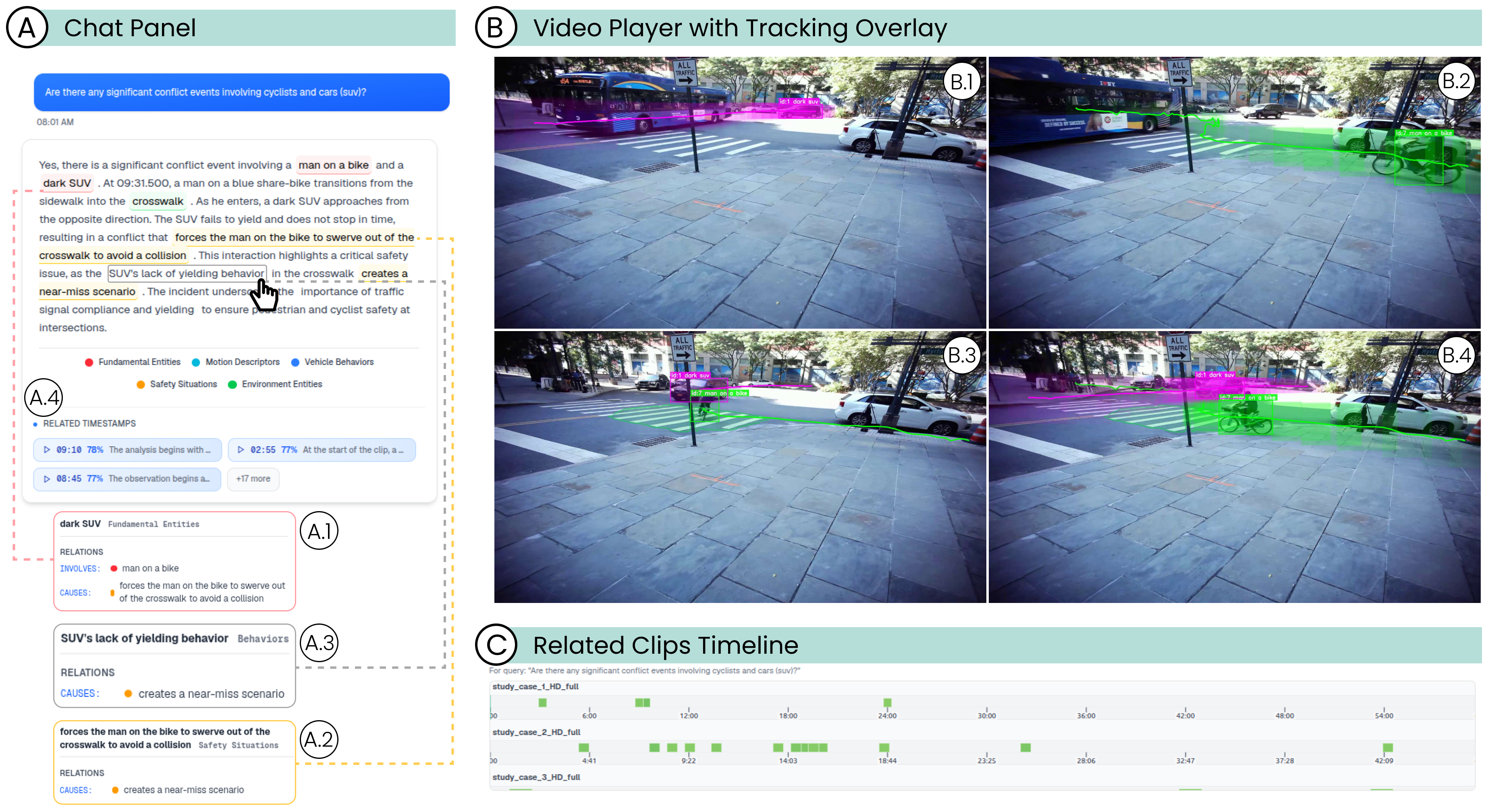}
  \vspace{-0.3cm}
 \caption{\textbf{\systemname interface.}
           (A) The \emph{Chat Panel} displays the user’s query, the RAG-generated narrative answer, and entity-level tooltips linked to the knowledge graph.
           (B) The \emph{Video Player with tracking overlays} shows the current frame with dynamic entities and highlighted static layout elements.
           (C) The \emph{Related Clips Timeline} summarizes retrieved clips across videos, with cells encoded by their semantic relevance to the query.
           }
    \label{fig:interface}
}

\maketitle
\begin{abstract}
Extracting actionable insights from long-duration urban videos is often labor-intensive: analysts must manually sift through raw footage to pinpoint target events or uncover broader behavioral trends.
In this work, we present \systemname, a visual analytics system for exploring long urban videos recorded at street intersections.
\systemname combines retrieval-augmented generation (RAG), taxonomy-aware entity extraction, and video grounding to support event retrieval and interpretation.
The system segments extended recordings into short clips, generates textual descriptions with a vision--language model, and indexes them for semantic retrieval.
A knowledge graph maps entities and relations from LLM answers onto a domain-specific taxonomy and aligns them with detected objects and trajectories to support visual grounding and verification.
\systemname supports scene retrieval through an augmented chat-based interface and improves scene interpretation by tightly aligning textual outputs with video evidence.
This design strengthens the connection between textual reasoning and visual evidence, reducing the effort required to validate model outputs and refine hypotheses.
We demonstrate the usefulness of \systemname on the StreetAware dataset through two case studies involving hazardous scenarios and crossing dynamics at street intersections.
\systemname helps analysts reason about safety- and mobility-related patterns across large urban video collections.


\end{abstract}  

\newcommand{\figInterface}{
\begin{figure*}[t!]
  \centering
  \includegraphics[width=0.99\linewidth]{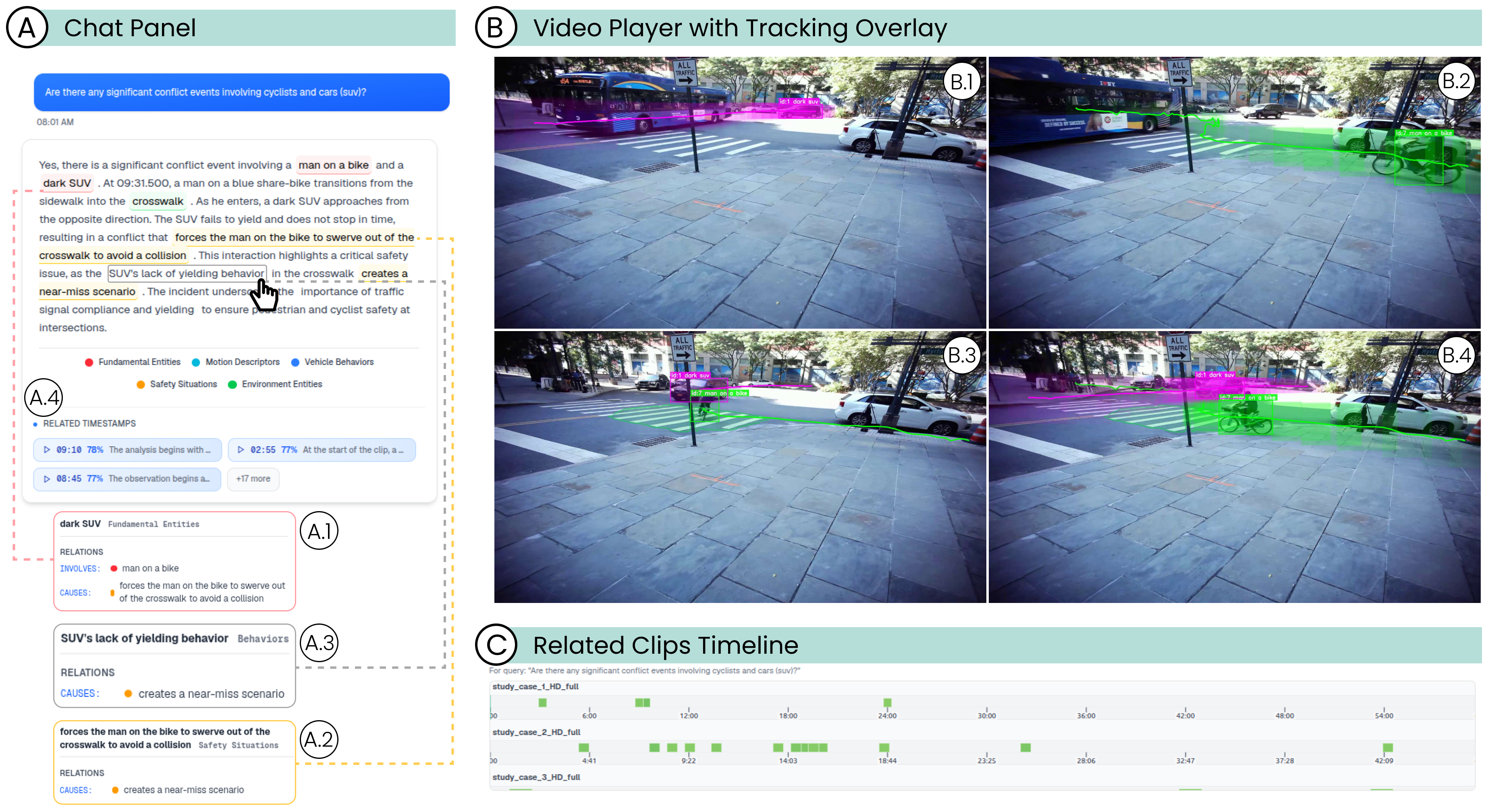}
  \caption{\textbf{\systemname interface.}
           (A) The \emph{Chat Panel} displays the user’s query, the RAG-generated narrative answer, and entity-level tooltips linked to the knowledge graph.
           (B) The \emph{Video Player with tracking overlays} shows the current frame with dynamic entities and highlighted static layout elements.
           (C) The \emph{Related Clips Timeline} summarizes retrieved clips across videos, with cells encoded by their semantic relevance to the query.
           }
    \label{fig:interface}
\end{figure*}
}
\newcommand{\figPipeline}{
\begin{figure*}[t!]
  \centering
  \includegraphics[width=\linewidth]{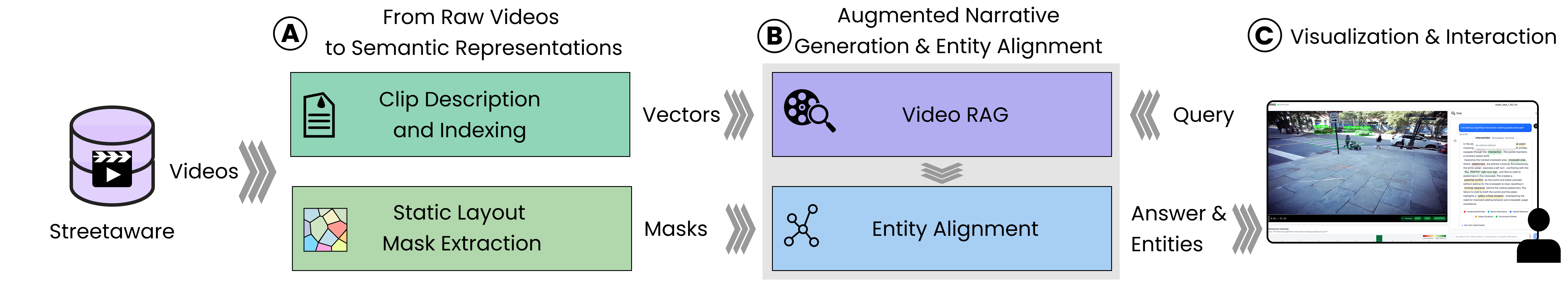}
  \caption{\textbf{\systemname main workflow.}
  (A) \emph{Preprocessing}: long videos are segmented into clips, described by a VLM, indexed in a vector store, and paired with static layout masks, yielding semantic and spatial representations for each video.
  (B) \emph{Augmented Narrative Generation}: at query time, the system enriches the user’s question, retrieves relevant clips, and composes narrative answers using the precomputed embeddings, knowledge graph, and masks.
  (C) \emph{Visualization}: the interface receives user queries, integrates outputs from all components, and supports interactive exploration and validation of retrieved content.}
  \label{fig:pipeline}
\end{figure*}
}

\newcommand{\figTaxo}{
\begin{figure*}[h!]
    \centering
    \includegraphics[width=1\linewidth]{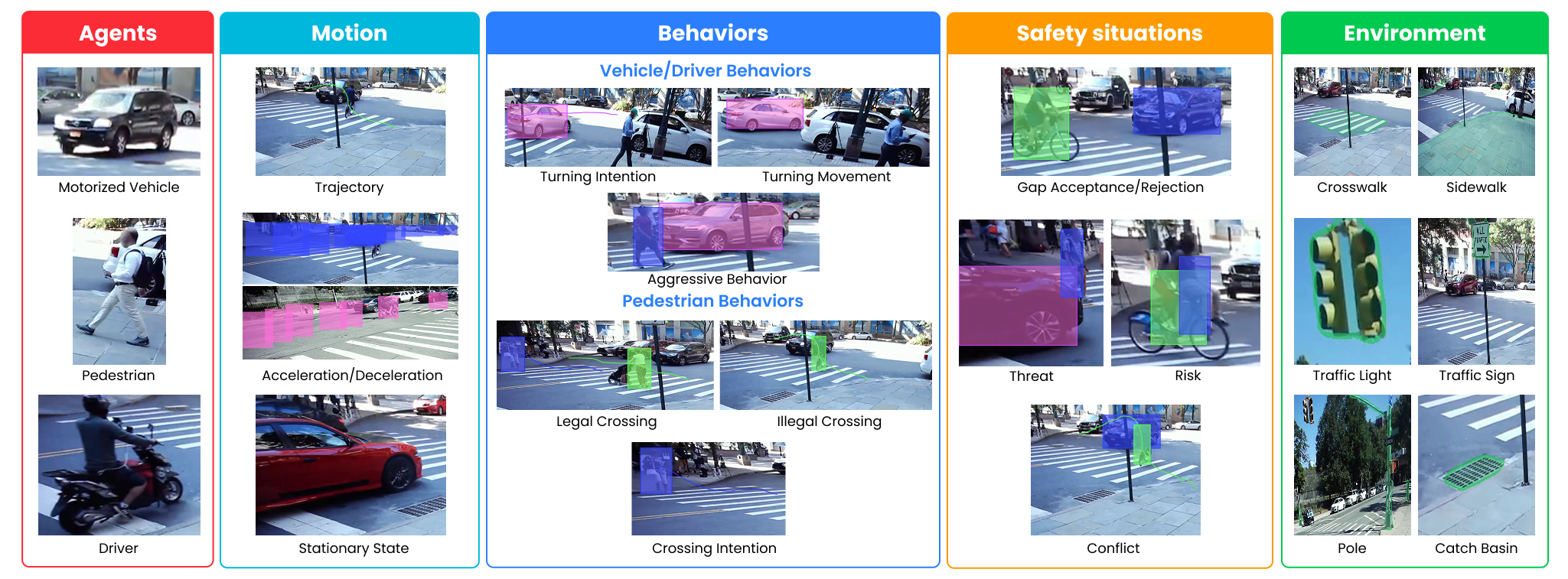}
    \caption{\textbf{Hierarchical taxonomy for knowledge-graph construction.}
    The taxonomy organizes urban-traffic concepts into five top-level categories---\emph{Agents}, \emph{Motion Descriptors}, \emph{Individual Behaviors}, \emph{Safety Situations}, and \emph{Environment Entities}---providing a consistent semantic basis for entity extraction, event classification, and grounding across all components of \systemname.}
    \label{fig:taxo}
\end{figure*}
}

\newcommand{\figAlign}{
\begin{figure}[t!]
    \centering
    \includegraphics[width=0.95\linewidth]{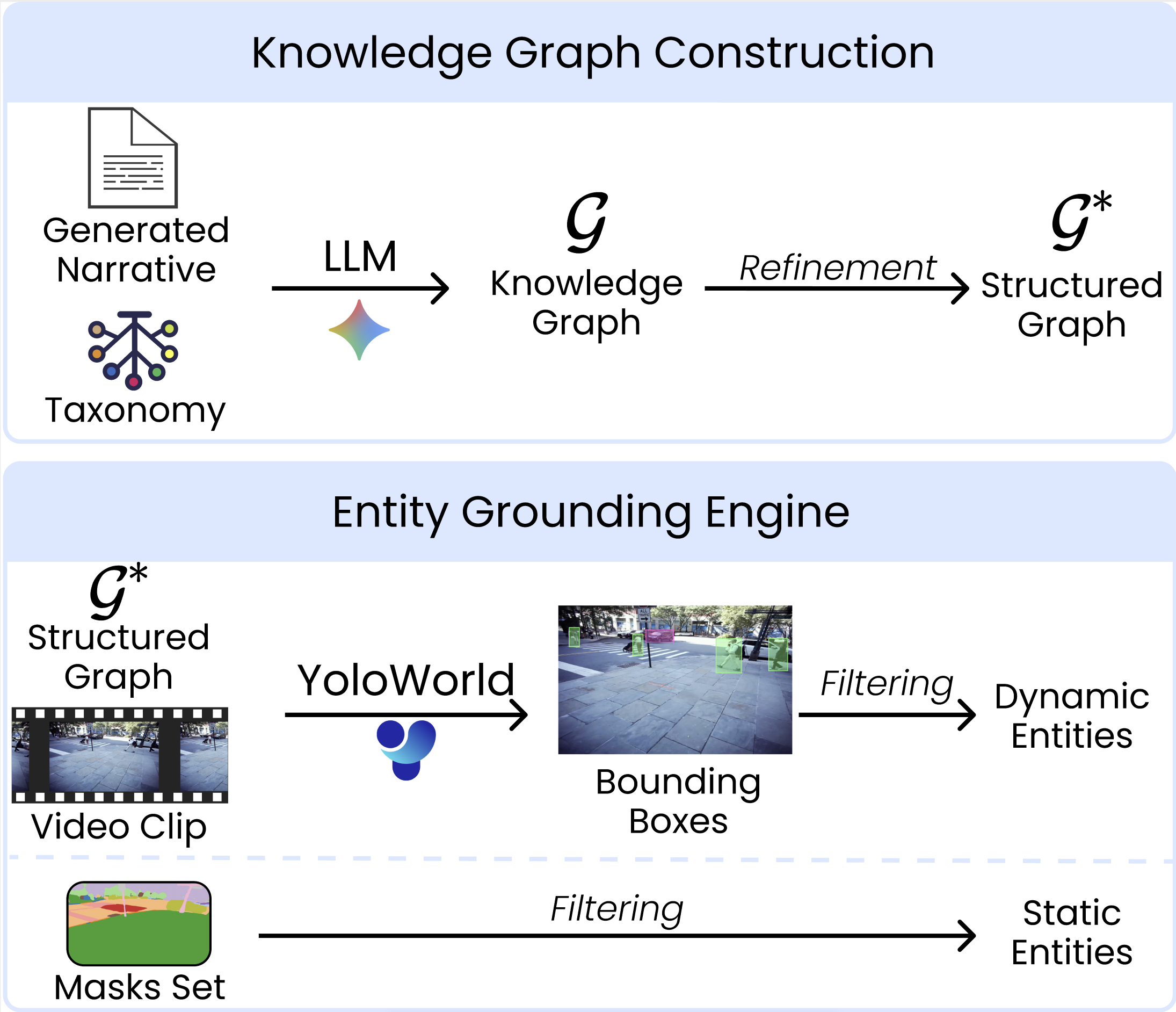}
    \caption{\textbf{Taxonomy-guided entity alignment.}
    The \emph{Knowledge Graph Construction} module uses the generated answer and the fixed taxonomy to build a structured graph $\mathcal{G}^*$ of entities and relations.
    The \emph{Entity Grounding Engine} then combines $\mathcal{G}^*$, the retrieved clip, and the precomputed masks to localize dynamic entities through detections and tracks, and to select relevant static environment masks for visualization.}
    \label{fig:ent_align}
\end{figure}
}

\newcommand{\figCaseOne}{
\begin{figure}[t!]
    \centering
    \includegraphics[width=0.95\linewidth]{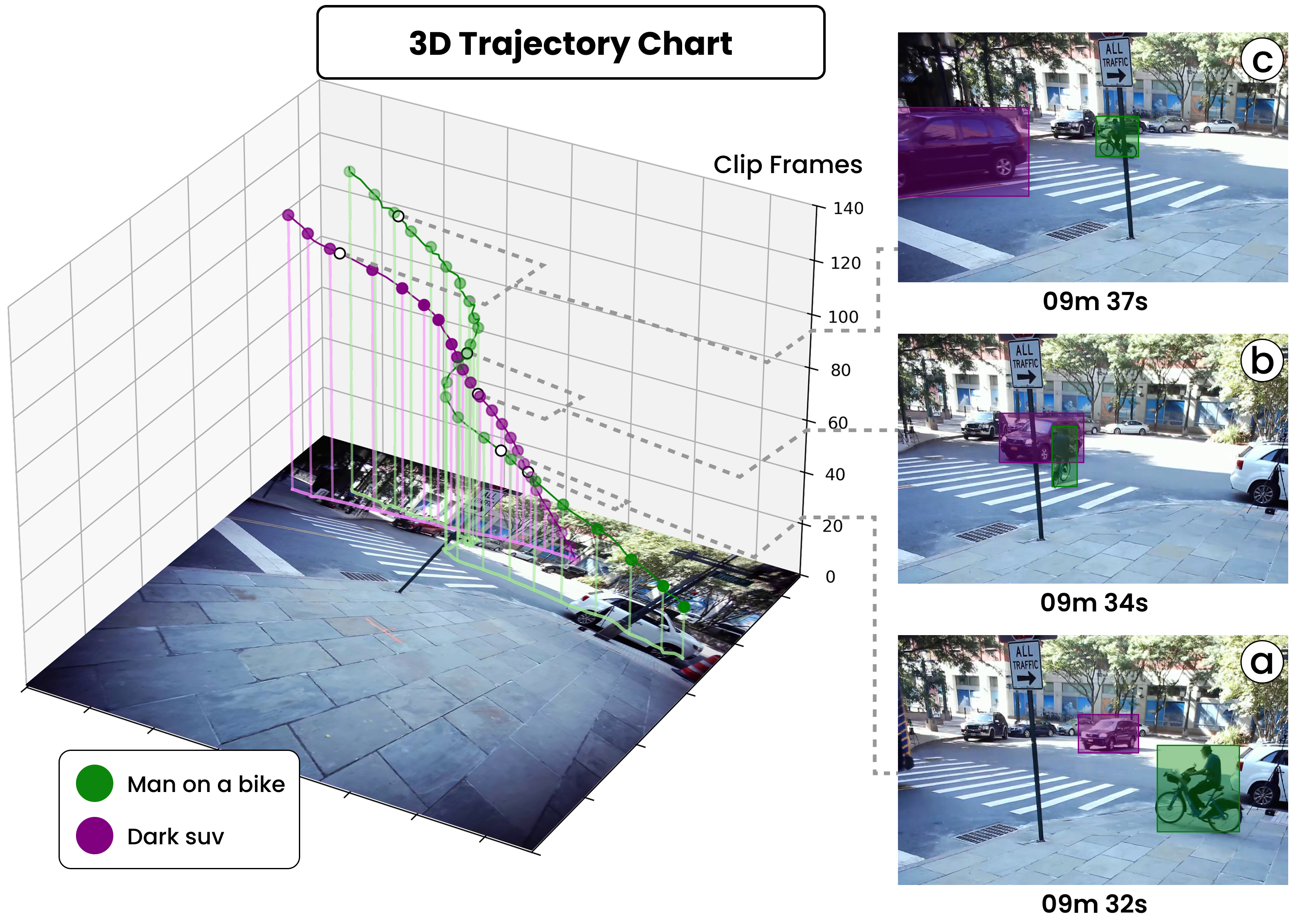}
    \caption{\textbf{Trajectories in Case Study~1.}
    The supporting 3D trajectory chart shows the motion of the \emph{man on a bike} (green) and the \emph{dark SUV} (purple) over time.
    Insets (a)--(c) show key frames: (a) the onset of the conflict as both approach the crosswalk; (b) the cyclist leaves the crosswalk while the SUV passes; and (c) the cyclist continues along the trajectory after the near miss.}
    \label{fig:suport_case1}
\end{figure}
}

\newcommand{\figCaseTwo}{
\begin{figure*}[h!]
  \centering
  \includegraphics[width=0.99\linewidth]{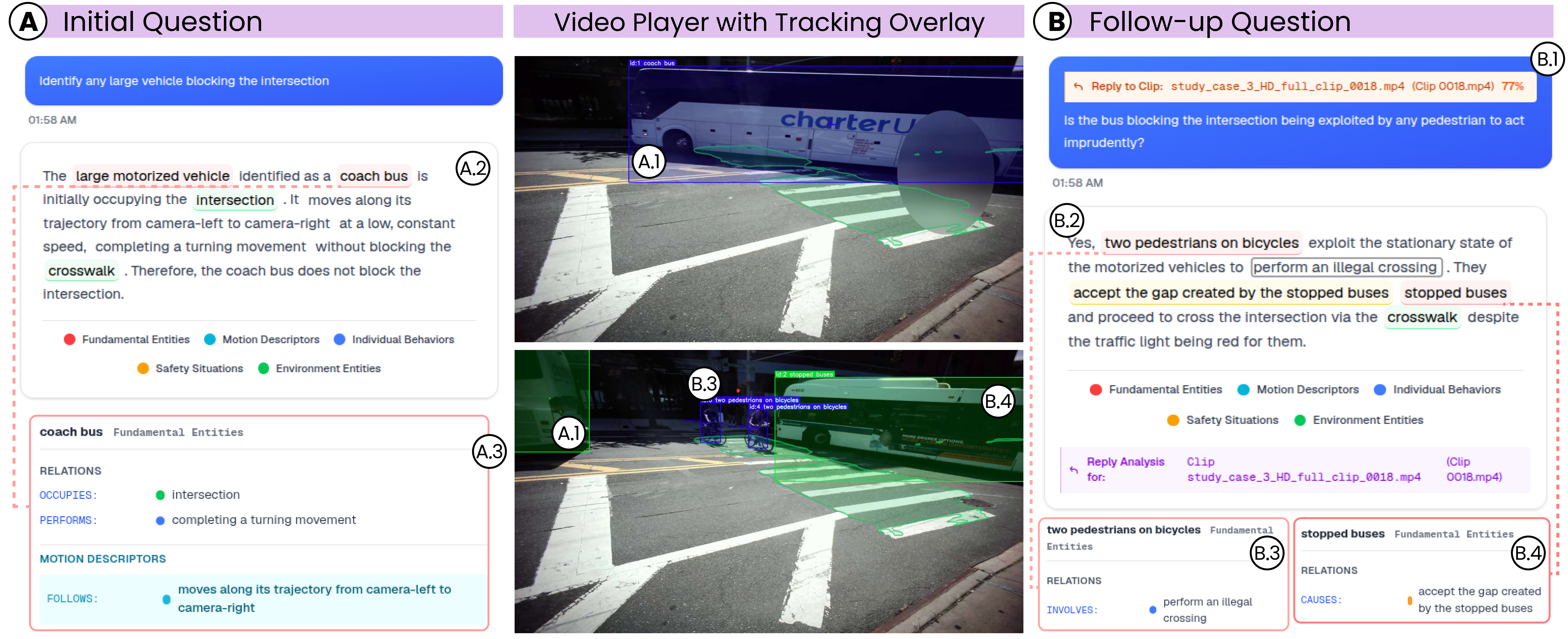}
  \caption{\textbf{Follow-up querying for risk-oriented analysis.}
    (A) Initial prompt asking for large vehicles blocking the intersection: the \textit{Chat Panel} (A.2--A.3) and \textit{Video Player} (A.1) highlight a coach bus executing a slow but compliant turning maneuver.
    (B) Follow-up prompt (B.1) explicitly focuses on safety risks associated with that scene: the updated narrative (B.2) and entity tooltips (B.3--B.4) reveal that a different bus enables an illegal crossing by pedestrians on bicycles.}
    \vspace{-0.5cm}
  \label{fig:study_case_2}
\end{figure*}
}
\section{Introduction}

Understanding the multifaceted dynamics of urban environments is fundamental to fostering resilient, safe, and equitable cities.
Street intersections are key nodes where diverse traffic flows converge, occupying only a small fraction of the street network yet playing a disproportionate role in safety and efficiency.
The National Association of City Transportation Officials (NACTO) calls them “one of the most critical and most complicated elements in roadway design,” and the US Department of Transportation (USDOT) reports that intersections account for a substantial share of traffic fatalities, with economic and societal costs reaching billions of dollars~\cite{FHWA_IntersectionSafety,blincoe2015economic}.

To optimize these locations, urban planners and transportation engineers rely on geometric and operational interventions—such as curb extensions, raised medians, and signal timing changes~\cite{stipancic2020pedestrian}.
Yet the intended effect of a design often differs from its real-world outcome.
Intersections are not just static infrastructure but dynamic environments shaped by human behavior, where pedestrians, cyclists, and vehicles negotiate shared space.
Assessing safety and efficiency thus requires granular observations of how these agents actually move and interact.

New sensing infrastructures now enable extended recording of intersection activity.
Unlike earlier approaches based on aerial imagery~\cite{chen2017surrogate,bahmanyar2025traffic} or GPS traces~\cite{deng2018generating,alkaissi2020delay}, modern systems capture rich, high-definition video directly from street level~\cite{piadyk_streetaware_2023}.
This street-view perspective provides nuanced information about vulnerable road users, fine-grained behavioral cues, and subtle interaction dynamics~\cite{rulff2024towards}.

Despite these advances, domain experts still lack scalable tools to retrieve and analyze specific events without laborious manual inspection.
Existing automated approaches typically rely on heavily supervised models and a limited vocabulary of predefined events~\cite{aboah2021vision,adewopo2024smart}, which restricts exploratory analysis and rapid adaptation to new policy questions. Still, new graph-LLM-based solutions are promising~\cite{2025-AnomalyExplain}.
Vision–language models (VLMs) offer a promising alternative: their zero-shot capabilities and natural-language interfaces allow analysts to describe scenarios in plain text and retrieve matching video segments.
However, interpreting VLM outputs in complex urban scenes remains challenging.
Analysts must relate long textual descriptions to visual counterparts involving many entities and intricate spatial relationships, a cognitively demanding process that is amplified when working with long-duration recordings.

There is thus a need for visual analytics techniques that structure and enrich VLM outputs, connect them to higher-level semantic models of intersection events, and ground these semantics in video to support sensemaking.
To address this need, we introduce \systemname, a visual analytics framework for event and scene retrieval in long-duration urban intersection videos.
\systemname{} combines a retrieval-augmented generation (RAG) architecture with taxonomy-aware entity extraction and smart video analysis to support both event retrieval and interpretation.
The system segments long recordings into clips, generates textual descriptions with a VLM, and indexes them for semantic retrieval.
At query time, natural-language questions are enriched to better match the indexed descriptions; relevant clips are retrieved; and entities mentioned in language-model responses are extracted and organized into a knowledge graph guided by a domain-specific taxonomy of intersection events.
Object detection and multi-object tracking then align these entities with visual instances, providing grounding that links semantic descriptions to video evidence.
In contrast to existing video RAG systems that operate mainly at the text-index level, our approach integrates retrieval, taxonomy-guided entity alignment, and video grounding within an interactive visual analytics framework.
This framework supports output-level auditing of black-box models through an augmented narrative view that presents model-generated answers alongside supporting video snippets, highlighted entities, and trajectories, reducing the effort required to interpret model outputs.
In summary, this work makes the following contributions:

\begin{itemize}
    \item \textbf{RAG-based video processing pipeline.}
    We propose a video RAG pipeline that segments long-duration recordings into clips, generates textual descriptions with a vision–language model, indexes them for semantic search, and performs prompt enrichment at query time to support interactive identification of salient video segments.
    
    \item \textbf{Taxonomy-aware entity alignment and visual grounding.}
    We introduce a knowledge-graph augmentation strategy that extracts entities from VLM responses, aligns them with detected visual objects, and organizes them according to a domain-specific taxonomy of urban intersection events, enabling consistent text--video grounding and higher-level reasoning.
    
    \item \textbf{Visual analytics system and evaluation.}
    We implement \systemname, an open-source visual analytics system that integrates these components into an augmented-narrative interface, bridging textual descriptions with visual evidence and supporting human-in-the-loop verification of model outputs.
    We demonstrate its usefulness through two case studies on the StreetAware dataset and report insights from structured interviews with urban specialists on usability and deployment potential.
\end{itemize}
\section{Related Work}

Our work lies at the intersection of RAG, video understanding, and visual analytics for urban monitoring.

\myparagraph{Retrieval-Augmented Generation for Video Understanding.}
RAG enhances the factual accuracy and knowledge scope of large language models (LLMs) by incorporating external knowledge sources~\cite{jeong_videorag_2025}.
Extending RAG to video introduces additional challenges, including the cost of converting long videos into text and the loss of multimodal cues such as motion and spatial context when relying solely on textual representations~\cite{ren_videorag_2025}.
These issues become particularly pronounced for long-duration corpora, where purely textual indexing often fails to preserve event-level structure~\cite{arefeen_irag_2024,ren_videorag_2025}.
To address these limitations, several systems adopt incremental pipelines that delay expensive processing until necessary.
Some enrich representations only for retrieved segments~\cite{arefeen_irag_2024}, while others maintain evolving knowledge structures to support low-latency retrieval as new content arrives~\cite{sankaradas_streamingrag_2025}.
Graph-based indices and hierarchical retrieval strategies~\cite{guo_lightrag_2025}, as well as cascaded pipelines combining lightweight models with VLM refinement~\cite{arefeen_vita_2024}, further reduce computational costs.

Another line of work strengthens retrieval by preserving multimodal signals rather than relying solely on text.
Approaches augment VLM inputs with auxiliary channels such as OCR, ASR, or object detections~\cite{luo_video-rag_2024}, apply question-aware frame sampling~\cite{tan_rag-adapter_2025}, or jointly reason across video, audio, and text~\cite{mao_multi-rag_2025}.
While these techniques improve retrieval relevance, they still struggle to capture long-range temporal dependencies.
More recent approaches construct structured representations across multiple videos, treating collections as interconnected knowledge spaces.
Graph-based grounding and multimodal encoders enable reasoning across entities and events at the corpus level~\cite{ren_videorag_2025,jeong_videorag_2025,trafficlens}.
However, these strategies introduce challenges related to entity canonicalization, graph maintenance, and multimodal alignment at scale.

Retrieval-oriented approaches are particularly promising for urban analytics, where scenes evolve rapidly and contextual grounding is essential~\cite{trafficlens,mao_multi-rag_2025}.
In contrast to these general-purpose retrieval methods, \systemname{} incorporates a domain-specific taxonomy and an explicit grounding loop to ensure that retrieved narratives remain aligned with the semantic requirements of urban safety analysis and can be audited against the underlying video evidence.

\myparagraph{Object Detection and Tracking for Grounding.}
Grounding language in video requires linking textual descriptions to entities in space and time through detection, tracking, and spatial reasoning.
Conventional closed-vocabulary detectors often struggle in open environments and fail to maintain consistent identities when objects are occluded or change appearance~\cite{jiang_comprehensive_2024}.
These issues are amplified in dense urban scenes involving heterogeneous agents.
Open-vocabulary detection addresses this limitation by aligning visual features with textual embeddings.
YOLO-World~\cite{cheng_yolo-world_2024} enables real-time detection of unseen categories, while OV-STAD~\cite{wu2024open} combines open-vocabulary detection with action understanding.
Despite these advances, maintaining temporally stable identities across frames remains challenging.

Recent work also emphasizes spatial reasoning.
Jiang et al.~\cite{jiang_comprehensive_2024} propose a pipeline that links entities, attributes, and relations across frames, supporting more temporally consistent alignments between visual content and generated text.
SpatialBot~\cite{cai2025spatialbot} leverages a VLM to reason about distances, directions, and geometric relations, enabling explicit spatial understanding.
Together, these systems point toward grounding models that can both identify objects and situate them in their environment.
These capabilities are vital for interpretable, context-aware video analysis, but they do not by themselves constrain how VLM or LLM outputs are used.
In \systemname, we close this loop by combining detection, multi-object tracking, and a specialized taxonomy into a grounding layer that explicitly checks model-generated claims against concrete visual evidence in urban intersection videos, exposing mismatches to the analyst.

\myparagraph{Visual Analytics for Video Data.}
Visual video analytics integrates computer vision with interactive visualization to support human reasoning over complex video data~\cite{schoning_visual_2019}.
Prior systems demonstrate the value of linking automated analysis to visual interfaces that enable analysts to inspect and guide algorithmic results.
Several systems combine visual features with textual information.
Wu and Qu~\cite{Wu2020multimodal} explore storytelling patterns in TED talks using transcripts and video features, while Chen et al.~\cite{Chen2020AVA} integrate trajectory analysis with spatial views to study traffic behavior.
Motion Browser~\cite{chan_motion_2020} similarly employs coordinated views to analyze motion patterns in clinical data.
Recent visual analytics systems have also explored interactive analysis of spatiotemporal phenomena through coordinated structural and temporal views, including association-rule-based exploration of categorical data and visual analysis of evolving geographic regions~\cite{diaz2025strive,nunes2025minetracker}.

Although these systems effectively expose spatiotemporal patterns, they typically rely on predefined features and offer limited support for natural-language queries and generative explanations.
Moreover, high-level descriptions are rarely grounded to concrete entities in the video.
To address these gaps, \systemname{} introduces a visual analytics framework that couples a RAG-based video pipeline with taxonomy-aware entity alignment and explicit grounding.
By anchoring model-generated explanations directly to retrieved clips and verifiable entities in the scene, the augmented narrative interface turns otherwise opaque VLM/LLM outputs into auditable, evidence-linked insights for urban safety analysis.

\figPipeline

\section{System Overview}

Prior work on RAG-based video retrieval and visual video analytics has highlighted persistent challenges in
(1) semantically searching long-duration videos,
(2) exposing entity-level spatiotemporal context, and
(3) supporting explanation and validation of algorithmic outputs directly in video frames
(\eg,~\cite{ren_videorag_2025,trafficlens,schoning_visual_2019,Wu2020multimodal,Chen2020AVA}).
Guided by these observations and formative discussions with urban traffic analysts, we derive a set of design goals and corresponding technical tasks for \systemname{}.
We then describe how these goals are operationalized through the three-stage workflow illustrated in Fig.~\ref{fig:pipeline}.

\subsection{Design Goals}

We conceptualize \systemname{} as a visual analytics system that sits between raw video streams and domain experts’ questions about safety and behavior at intersections.
The following design goals capture the capabilities needed to analyze large collections of urban videos.

\begin{description}
    \item[G1] \textbf{Text-based semantic access to long-duration videos.}
    Enable analysts to formulate information needs in natural language and retrieve semantically relevant video segments, avoiding manual inspection of hours of footage.

    \item[G2] \textbf{Entity-centric spatiotemporal context.}
    Provide representations that expose the locations of entities (\eg, vehicles and pedestrians) in the scene and how their states and interactions evolve over time, supporting reasoning about trajectories, conflicts, and usage patterns.

    \item[G3] \textbf{Transparent, visually grounded explanations.}
    Couple model-generated narratives with explicit visual evidence so that analysts can quickly understand, verify, and critique system outputs rather than treating them as black-box predictions, thereby supporting output-level auditing of VLM/LLM behavior in real scenes.
\end{description}

\subsection{Design Tasks}

To operationalize these goals in a visual analytics setting, we specify the following technical design tasks for \systemname:

\begin{description}
    \item[T1] \textbf{Natural-language query parsing and semantic retrieval (G1).}
    Implement a text interface that accepts free-form questions, performs query enrichment with domain terminology, encodes text into an embedding space, and retrieves candidate clips from a vector index of video descriptions.

    \item[T2] \textbf{Entity extraction, taxonomy mapping, and knowledge-graph construction (G2, G3).}
    From RAG-generated answers, automatically detect entity mentions such as a car turning right or a pedestrian crossing late, map these mentions to a domain-specific taxonomy of intersection events, and instantiate a knowledge graph that captures entity types, roles, and relationships.

    \item[T3] \textbf{Spatiotemporal grounding via detection and tracking (G2, G3).}
    Run object detection and multi-object tracking over video clips, associate tracked objects with textual entities from the knowledge graph, and maintain persistent identifiers and trajectories across frames and clips.

    \item[T4] \textbf{Linked visual encodings for narrative, entities, and video (G2, G3).}
    Design coordinated views so that the augmented narrative, entity list/graph, and video player share common references. Selecting an entity in the text highlights the corresponding track in the video and vice versa.

    \item[T5] \textbf{Interaction techniques for clip navigation and comparison (G1–G3).}
    Provide mechanisms to navigate between retrieved clips, refine queries, and compare alternative narratives (\eg, multiple queries or parameter settings), supporting iterative exploration of complex behaviors across the video corpus and human-in-the-loop refinement of the system’s interpretations.
\end{description}

\subsection{System Workflow}

To realize these goals, \systemname{} transforms raw urban videos into an interactive analysis environment through three integrated stages (Fig.~\ref{fig:pipeline}).
In the \emph{Preprocessing} stage (A), the system segments long-duration street-level recordings, such as those from the StreetAware dataset~\cite{piadyk_streetaware_2023}, into overlapping clips.
It then processes these clips offline to obtain two complementary representations: textual descriptions generated by a vision-language model and stored in a vector index for semantic retrieval, and static layout masks that describe the intersection infrastructure, such as crosswalks, sidewalks, and lanes.
During \emph{Augmented Narrative Generation} (B), analysts submit natural-language questions.
\systemname{} enriches these queries with domain terminology, retrieves relevant clips from the semantic index, and uses a RAG pipeline to compose narrative answers.
Entities are extracted and mapped to a taxonomy of intersection events, and then instantiated in a knowledge graph that can be aligned with downstream object detection and tracking results.
Finally, the \emph{Visualization and Interaction} stage (C) presents these results through an augmented narrative interface that coordinates a video player, a narrative view, and an entity/graph view.
Analysts can move smoothly between queries, explanations, and visual evidence to explore and validate safety- and mobility-related patterns across the video corpus.

\section{From Raw Videos to Semantic Representations}

To support semantic search and retrieval over long urban videos, \systemname performs an offline preprocessing stage (Fig.~\ref{fig:pipeline}A) that converts raw intersection footage into structured, queryable representations.
This stage comprises two complementary procedures: (1) clip description and indexing for text-based retrieval and (2) static layout mask extraction to contextualize events within the physical infrastructure.

\figTaxo

\subsection{Clip Description and Indexing}

Each input video $v$ has duration $T_v$ seconds (not necessarily the same across videos).
To preserve continuity and reduce the risk of missing important interactions at segment boundaries, we divide each video into fixed-length clips ($\tau = 30$ seconds) with a $5$-second overlap ($\omega = 5$).
For a given video $v$, the number of clips is
\[
    m_v = \left\lceil \frac{T_v - \tau}{\tau - \omega} \right\rceil + 1,
\]
and the $i$-th clip $C_{v,i}$ starts at timestamp
\(
    t_{v,i} = (i - 1)\,(\tau - \omega).
\)

Each segmented clip $C_{v,i}$ is then processed by a VLM to generate a textual description summarizing its visual content and activities.
Given a prompt template $\mathcal{P}$ (Supplementary Material C.1), designed to emphasize traffic entities, behaviors, and safety-critical situations, we obtain $D_{v,i} = \mathrm{VLM}(C_{v,i}, \mathcal{P}).$
During this stage, we also adjust timestamps when needed to ensure temporal consistency across descriptions by correcting offsets introduced by segmentation or buffering.

To enable semantic similarity search, the textual descriptions are embedded in a continuous vector space.
An embedding model $\mathrm{E}$ encodes each description $D_{v,i}$ as
\[
    \mathbf{e}_{v,i} = \mathrm{E}(D_{v,i}), \quad \mathbf{e}_{v,i} \in \mathbb{R}^d,
\]
where $d$ denotes the dimensionality of the embedding space.
We store these embeddings, together with their metadata (timestamps, clip identifiers, and original descriptions), in a vector database, yielding an indexed collection of clip descriptors that supports efficient semantic retrieval during the query phase.

\subsection{Static Layout Mask Extraction}
\label{sec:videomask}
To provide spatial context for retrieved events and to distinguish moving agents from static infrastructure, \systemname extracts semantic masks of each intersection's layout.
The goal is to obtain a clean representation of the environment (\eg, crosswalks, sidewalks, and lanes).

For each video $v$, we first sample $n$ frames uniformly across its duration, denoted $F_{v,1}, \dots, F_{v,n}$.
Each sampled frame is processed with the open-vocabulary detector YOLO-World~\cite{cheng_yolo-world_2024} to identify dynamic entities such as vehicles, pedestrians, and cyclists.
We selected YOLO-World for its state-of-the-art open-vocabulary detection, which allows us to specify the traffic-related categories of interest.
To reduce false positives and enhance precision, we apply a detection confidence threshold of 0.65 across all YOLO-World outputs, based on empirical tuning and used consistently in all our experiments.

\looseness=-1
Let $N(F_{v,j})$ denote the number of detected objects in frame $F_{v,j}$.
We choose as a reference the frame with the lowest detected object count,
\(
    F_v^* = \arg\min_{j \in \{1,\dots,n\}} N(F_{v,j}),
\)
which in practice yields a view of the intersection where static infrastructure is least occluded by traffic.
We then apply the semantic segmentation model Mask2Former~\cite{cheng2022masked} to $F_v^*$ to obtain a set of masks $\mathcal{M}_v = \{M_{v,1}, \dots, M_{v,p}\},$
where each mask corresponds to a specific static component of the environment (\eg, crosswalks, traffic signals, road surfaces, sidewalks, and lane markings).
These layout masks serve as spatial reference layers throughout the pipeline, enabling subsequent modules to contextualize detected events relative to the intersection's physical geometry and supporting later tasks such as entity-level grounding and spatial reasoning.

\section{Augmented Narrative Generation and Entity Alignment}

The answer generation stage is the core of \systemname (Fig.~\ref{fig:pipeline}B).
It operates on the preprocessed clip index and layout masks and (i) retrieves relevant clips through a video RAG pipeline and (ii) structures and grounds the resulting narrative using a taxonomy-guided knowledge graph.

\subsection{Video RAG}
\label{sec:videorag}

The video RAG component processes a user’s natural-language query through three steps: query enrichment, semantic retrieval, and narrative generation.
It produces (1) a query-aligned narrative answer and (2) metadata about the selected segments as evidence.


\myparagraph{Query Enrichment.}
Given a user query $Q$, we first expand it into an enriched query $Q'$ to better match the indexed clip descriptions.
A language model augments $Q$ with traffic-specific terminology, paraphrases, and contextual hints drawn from a domain context $\mathcal{K}$ (\eg, intersection events and entities) as $Q' = \mathrm{LLM}(Q, \mathcal{K}).$
This improves recall by reducing vocabulary mismatch and also helps filter invalid or out-of-domain questions.

\myparagraph{Semantic Retrieval.}
We embed the enriched query $Q'$ in the same semantic space as the clip descriptions defined in the preprocessing stage using the embedding model $\mathrm{E}$:
\[
    \mathbf{q} = \mathrm{E}(Q'), \quad \mathbf{q} \in \mathbb{R}^d.
\]
The vector database then computes cosine similarity between $\mathbf{q}$ and all clip embeddings $\mathbf{e}_{v,i}$, ranks clips by similarity, and returns the top-$k$ candidates $\mathcal{R}_k = \{(v,i)\}_{j=1}^{k},$
where each pair $(v,i)$ indexes a retrieved $C_{v,i}$ and its associated $D_{v,i}$.

\myparagraph{Narrative Generation.}
The narrative answer is generated by conditioning an LLM on the original query and the retrieved evidence.
In our implementation, we use the highest-ranked clip $C_{v,i}$ together with its description and temporal metadata $ A = \mathrm{GenerateNarrative}(Q, C_{v,i}, D_{v,i}, t_{v,i}) $.
Where $A$ follows domain terminology and explicitly references intersection entities, behaviors, and safety situations.
In the next stage, we convert $A$ into a structured representation and align its entities with the video content.

\subsection{Taxonomy-Guided Entity Alignment}
\label{sec:entity-alignment}

Raw answers can be long and difficult to interpret.
To help analysts understand and validate the generated narratives, \systemname augments them with a structured, grounded representation.
This process has two parts—knowledge graph construction and entity grounding—whose main components are summarized in Fig.~\ref{fig:ent_align}.
The top panel depicts the Knowledge Graph Construction module, and the bottom panel depicts the Entity Grounding Engine.

\myparagraph{Knowledge Graph Construction.}
We represent the answer as a knowledge graph $\mathcal{G} = (\mathcal{V}, \mathcal{E})$, where nodes $\mathcal{V}$ denote entities and edges $\mathcal{E}$ denote relationships.
The \emph{Knowledge Graph Construction} module (top part of Fig.~\ref{fig:ent_align}) performs this transformation.
Entity extraction is guided by a fixed crossroad-interaction taxonomy $\mathcal{T}$ inspired by Shirazi and Morris~\cite{shirazi_looking_2017} and adapted to our setting.
Fig.~\ref{fig:taxo} illustrates this taxonomy, which organizes concepts into five top-level categories: agents, motion descriptors, individual behaviors, safety situations, and environment entities, together with examples of each entity type considered.

An LLM-based extractor takes the answer $A$ and outputs entity mentions.
Each mention has a class in $\mathcal{T}$, a text span, optional attributes (\eg, \emph{red car}, \emph{turning right}), and a position in the answer.
We then perform a light canonicalization step to merge mentions of the same real-world entity (\eg, “a red car” and later “the vehicle” in the same scene) using lexical overlap and LLM-based coreference scoring.
The result is a deduplicated set of nodes $\mathcal{V}^*$ with enriched attributes.

Relations are extracted as directed triples $(u_s, r, u_o)$ where $u_s, u_o \in \mathcal{V}^*$ and $r$ is a relation type such as \emph{approaches}, \emph{yields-to}, or \emph{conflicts-with}.
The resulting graph $\mathcal{G}^* = (\mathcal{V}^*, \mathcal{E})$ provides a compact semantic summary of the narrative that can be visualized and used for grounding.


\figAlign

\myparagraph{Entity Grounding Engine.}
While $\mathcal{G}^*$ captures the semantics of the answer, analysts also need to see where entities appear in the video.
The Entity Grounding Engine (bottom part of Fig.~\ref{fig:ent_align}) attaches visual footprints (bounding boxes or masks) to the entities in $\mathcal{V}^*$ by combining YOLO-World detections with the static layout masks from Sec.~\ref{sec:videomask}.

For \emph{dynamic entities} (\eg, vehicles and pedestrians), we use YOLO-World with the textual description of each relevant entity $u \in \mathcal{V}^*$ as prompt.
Given a frame $F_t$ at time $t$, the detector returns a set of bounding boxes $\mathcal{B}_u = \mathrm{YOLOWorld}(F_t, \text{span}(u)),$
where each $b \in \mathcal{B}_u$ encodes position, size, and confidence.
Across frames, these detections can be linked into short tracks.

For \emph{static environment entities} (\eg, crosswalks, signals, and lane markings), grounding uses the precomputed layout masks $\mathcal{M}_v$ for the corresponding video.
We select only those masks whose class appears in the entity set: $\mathcal{M}_{\text{active}} = \{M \in \mathcal{M}_v \mid \mathrm{class}(M) \in \mathrm{classes}(\mathcal{V}^*)\}.$
Restricting to $\mathcal{M}_{\text{active}}$ avoids visual clutter and focuses attention on environmental elements relevant to the current answer.

Together, the knowledge graph and grounding engine connect textual explanations to concrete visual evidence, enabling the interface to highlight entities in both the narrative and the video and to support entity-centric exploration of complex intersection scenes.
\section{\systemname}

The \systemname interface (Fig.~\ref{fig:interface}) is designed as an interactive visual analytics environment centered on an augmented conversational experience.
It supports the exploration, retrieval, and validation of specific events in long-duration urban videos by tightly coupling narrative answers, entity-level structures, and video evidence.
The interface is organized into three coordinated components: the Chat Panel, the Video Player with tracking overlays, and the Related Clips Timeline.


\subsection{Chat Panel}
\label{sec:querypanel}

The \textit{Chat Panel} (Fig.~\ref{fig:interface}A) is the main entry point for analysts.
It receives the user’s natural-language query and displays the augmented narrative answer $A$ produced by the video RAG pipeline.
The answer describes what happens in the retrieved clip and identifies key actors (\eg, pedestrians, vehicles, and cyclists) as well as environmental elements (\eg, crosswalks, traffic lights, and sidewalks).
Using the entity metadata from the knowledge graph, the panel highlights mentions of entities directly in the text.
Entities are color-coded according to the taxonomy categories introduced in Fig.~\ref{fig:taxo}, with a legend shown at the bottom of the panel (Fig.~\ref{fig:interface}A.4).
Clicking or hovering over a highlighted entity links its mention in the text to the corresponding visualization in the video, allowing analysts to directly identify and confirm the entity's role and actions within the event.

When hovering over an entity, an interactive tooltip appears (Fig.~\ref{fig:interface}A.1--A.3).
This tooltip summarizes the local neighborhood of that node in the knowledge graph, showing related entities and relationships (\eg, \emph{involves} and \emph{causes}).
This design follows validation strategies in prior visual analytics systems that couple LLM reasoning with graph structures~\cite{coscia_vispile_nodate}, helping users assess how the model connects actors, behaviors, and safety situations.

The \textit{Chat Panel} also surfaces links to additional clips that exhibit similar behaviors or interactions (Fig.~\ref{fig:interface}A.4), together with a confidence indicator reflecting the relevance of the retrieved clip to the user’s query.
Selecting one of these related clips updates the \textit{Video Player} and the \textit{Related Clips Timeline}.
This supports comparison of how a given pattern manifests at different times or in different locations.

\subsection{\textit{Video Player} with Tracking Overlays}
\label{sec:videoplayer}

The \textit{Video Player} (Fig.~\ref{fig:interface}B) provides the primary visual evidence supporting the narrative answer.
Its role is to allow analysts to visually confirm and contextualize the retrieved information.
The textual explanation, the knowledge graph, and the video frames remain tightly aligned.
To support this, the player renders multiple visual layers on top of each frame in response to interactions originating in the Chat Panel or the Timeline.

\myparagraph{Dynamic entities.}
Dynamic actors, such as vehicles and pedestrians, are localized using YOLO-World.
This model is prompted with textual descriptions of entities extracted from the answer.
They are rendered as colored bounding boxes whose hues follow the taxonomy.
Examples of these overlays are shown in Fig.~\ref{fig:interface}B.1, B.2, and B.4.
Clicking a dynamic entity mention in the \textit{Chat Panel} jumps the video to the first frame in which that entity is detected and displays its subsequent trajectory.

\myparagraph{Static environment entities.}
Static infrastructure elements (\eg, crosswalks, sidewalks, lanes, and poles) are rendered using the precomputed layout masks from Sec.~\ref{sec:videomask}.
These masks are drawn more subtly than dynamic entities to provide spatial context without overwhelming the scene (Fig.~\ref{fig:interface}B.3 and B.4).
Hovering over an environmental entity in the text highlights the corresponding region in the video.
This interaction makes explicit how narrative elements relate to physical infrastructure, revealing where and how entities interact with the environment.

All overlays remain synchronized with temporal navigation.
Scrubbing, playing back, or jumping via the Timeline keeps the highlighted entities and trajectories aligned with the narrative.
This enables analysts to move seamlessly between detailed frame-level inspection and higher-level reasoning about behaviors and safety situations.

\subsection{Related Clips Timeline}
\label{sec:tempheat}

The \textit{Related Clips Timeline} (Fig.~\ref{fig:interface}C) acts as a temporal overview and navigation aid for the retrieved results.
It visualizes clips ranked by semantic relevance to the current query, as computed by the video RAG component (Sec.~\ref{sec:videorag}).

Each row in the timeline corresponds to a video, and each cell corresponds to a retrieved clip within that video.
The horizontal position encodes the clip’s temporal location in the source video, while cell appearance encodes its semantic similarity score (clips more relevant to the query are shown more prominently).
When the user restricts the search to a single video, the timeline displays a single row.
When the search spans the corpus, multiple rows appear, enabling analysts to compare how the queried event or behavior manifests across different viewpoints or intersections.

Interaction with the timeline is tightly coupled to the other views.
Clicking a cell in the active video row causes the \textit{Video Player} to seek to the start of that clip.
Selecting a cell from another video switches the player to that source, updates the \textit{Chat Panel}, and updates the overlays accordingly.
This coordination allows analysts to move smoothly between corpus-level exploration (identifying where relevant events occur) and detailed inspection of individual scenarios.

\subsection{Implementation Details}

\systemname\ is implemented as a Python backend using \texttt{FastAPI} for HTTP and WebSocket communication and a \texttt{Svelte + TypeScript} frontend with \texttt{D3.js} for interactive visualizations.
The system integrates multiple foundation models, including \texttt{gemini-2.5-pro} for video captioning, \texttt{gemini-embedding-001} for semantic indexing, and \texttt{gpt-4o} for query enrichment and narrative generation.
Object grounding is performed with YOLO-World, while static layout masks are extracted using Mask2Former.

At query time, the end-to-end response latency is typically around 6--7 seconds on a consumer workstation.
Additional implementation details, system configuration, and prompts used in the LLM pipeline are provided in Supplementary Material ~A and ~C.
To support reproducibility, the source code, configuration files, and documentation for \systemname\ are publicly available at \url{https://visualdslab.com/papers/UrbanClipAtlas/}.
\figCaseOne

\section{Evaluation}

We qualitatively evaluate \systemname using the \textit{StreetAware} dataset and two case studies.
The goal is to illustrate how the system supports event retrieval, multi-entity reasoning, and visual validation of safety-related situations through concrete examples based on real urban videos.

\subsection{Dataset: StreetAware}

We use the \textit{StreetAware} dataset~\cite{piadyk_streetaware_2023}, which contains high-resolution, synchronized, multimodal street-level videos.
These videos were captured at busy urban intersections, with privacy preserved through the removal of audio and the blurring of vehicle license plates and human faces.
StreetAware focuses on high-activity crossings in Brooklyn, New York, and includes approximately 8 hours of recordings from three intersections.
Unlike traditional automotive datasets captured from a moving ego-vehicle (\eg, KITTI and Cityscapes), StreetAware employs static, multi-angle cameras mounted at intersection corners.
This setup provides a stable, wide view of pedestrian--vehicle interactions and the surrounding built environment.
It is therefore well suited for analyzing behaviors such as yielding, jaywalking, and near-miss incidents.

\figCaseTwo
\vspace{-5pt}
\subsection{Case Studies}

We present two case studies to demonstrate \systemname's capabilities.
Each case study starts from a natural-language query and shows how the \textit{Chat Panel}, \textit{Video Player}, and \textit{Related Clips Timeline} (Fig.~\ref{fig:interface}) work together to surface relevant events and support detailed inspection.

\myparagraph{Conflicts between cyclists and cars.}
In our taxonomy (Fig.~\ref{fig:taxo}), a \textit{Conflict} is a \textit{Safety Situation} involving two or more dynamic entities whose trajectories bring them into spatial and temporal proximity, creating a risk of collision without an actual crash~\cite{shirazi_looking_2017}.
This case study examines how \systemname helps identify and analyze such conflicts between cyclists and vehicles.
The analyst issues the query:
\emph{“Are there any significant conflict events involving cyclists and cars (SUV)?”}
The Video RAG component retrieves a set of relevant clips and generates a narrative answer, which appears in the \textit{Chat Panel} (Fig.~\ref{fig:interface}A).
The answer highlights three key entities: \emph{dark SUV}, \emph{man on a bike}, and \emph{crosswalk}, and describes their interactions.
The knowledge-graph tooltips (Fig.~\ref{fig:interface}A.1--A.3) expose the main relations:
\begin{itemize}
    \item \textit{dark SUV} $\xrightarrow{\textit{involves}}$ \textit{man on a bike};
    \item \textit{dark SUV} $\xrightarrow{\textit{causes}}$ \textit{forces the man on the bike to swerve out of the crosswalk};
    \item \textit{forces the man on the bike to swerve out of the crosswalk} $\xrightarrow{\textit{causes}}$ \textit{creates a near-miss scenario}.
\end{itemize}
These relations collectively characterize the situation as a conflict event.
The \textit{Video Player} with tracking overlays (Fig.~\ref{fig:interface}B) provides the corresponding visual evidence.
The dark SUV and the cyclist are rendered as colored tracks, allowing the analyst to observe how the cyclist changes course to avoid the vehicle.
Clicking the relevant entities in the \textit{Chat Panel} centers the playback on the moment of maximum risk and reveals the full trajectories of both actors.

To further support reasoning about motion, we use the 3D trajectory view (an external visualization not implemented directly within \systemname) shown in Fig.~\ref{fig:suport_case1}.
The ground plane shows the reference frame of the intersection, while the vertical axis encodes time.
The green trajectory corresponds to the \emph{man on a bike}, and the purple one to the \emph{dark SUV}.
Panels (a)--(c) on the right show key frames: (a) when the conflict begins as both approach the crosswalk; (b) the cyclist leaves the crosswalk while the SUV passes; and (c) the cyclist resumes their trajectory.
Although the projected paths intersect on the ground plane, the temporal axis reveals that the two entities do not occupy the same space at the same time.
This confirms the event as a near miss rather than a collision.

The \textit{Related Clips Timeline} (Fig.~\ref{fig:interface}C) lists additional clips ranked by semantic similarity to the query.
Most retrieved clips depict interactions between cyclists and vehicles at varying levels of risk.

\noindent\textit{Take-away.}
\systemname successfully retrieves conflict events that match the analyst’s query and explains them through coordinated narrative, structure, and video.
The \textit{Chat Panel} and knowledge graph clarify which entities and relations define the conflict, while the \textit{Video Player}, the trajectory visualization, and the \textit{Related Clips Timeline} support inspection of how the event unfolds and how it relates to similar scenarios.
Together, these components support both the discovery of near-miss events and the nuanced interpretation of their dynamics.

\myparagraph{Illegal crossing triggered by bus occlusion.}
A key advantage of a chat-based interface is that analysts can iteratively refine their questions, moving from broad descriptions to targeted risk assessments.
In this case study, we start with a general request:
\emph{“Identify any large vehicle blocking the intersection.”} (Fig.~\ref{fig:study_case_2}A.1).
We chose this example because multiple large vehicles traverse the intersection throughout the day.
In response, \systemname retrieves a clip in which a coach bus initially occupies part of the intersection.
The \textit{Video Player} (Fig.~\ref{fig:study_case_2}A.1) highlights the bus with a bounding box, and the narrative in the \textit{Chat Panel} explains its behavior:
\emph{“The coach bus is initially occupying the intersection. It moves along its trajectory from camera-left to camera-right at a low, constant speed, completing a turning movement without blocking the crosswalk.”}
This description, together with the entity-level breakdown in the tooltip (Fig.~\ref{fig:study_case_2}A.2--A.3), indicates that the bus clears the crosswalk and the intersection without creating an obstruction.
At this stage, the system characterizes the vehicle’s behavior as compliant.

However, slow turning movements by large vehicles can still create risky situations, particularly when other road users attempt to exploit temporary gaps.
This concern motivates a follow-up question that focuses explicitly on risk:
\emph{“Is the bus blocking the intersection being exploited by any pedestrian to act imprudently?”} (Fig.~\ref{fig:study_case_2}B.1).
Since the query builds on the previous clip, the system maintains context while refocusing the narrative toward safety implications.
The new answer (Fig.~\ref{fig:study_case_2}B.2) reveals an emergent pattern:
\emph{“Yes, two pedestrians on bicycles exploit the stationary state of the motorized vehicles to perform an illegal crossing. They accept the gap created by the stopped buses and proceed to cross the intersection via the crosswalk despite the traffic light being red for them.”}
Entity extraction and the knowledge-graph tooltips (Fig.~\ref{fig:study_case_2}B.3--B.4) make this structure explicit by connecting \textbf{two pedestrians on bicycles} to the action \textbf{perform an illegal crossing} and linking \textbf{stopped buses} to the relation \textbf{causes: accept the gap created by the stopped buses}.
In other words, the buses form a visual shield that encourages risky behavior at the crosswalk.

Importantly, the bus implicated in this second narrative is not the original coach bus from the initial query (Fig.~\ref{fig:study_case_2}A.1), but a different bus that arrives immediately afterward (Fig.~\ref{fig:study_case_2}B.4).
This follow-up reframes the scene: the initial query shows compliant bus behavior, but the refined question reveals how a subsequent bus creates conditions for an illegal red-light crossing.

\noindent\textit{Take-away.}
This case study illustrates how \systemname supports semantic retrieval and iterative, risk-focused exploration of long-duration urban videos.
By allowing follow-up questions anchored to specific clips, the system helps analysts move from verifying compliant vehicle behavior to uncovering subtle safety risks.
Here, an illegal crossing emerges due to bus occlusion, and analysts can identify it without manually re-scrubbing the footage.
\section{Discussion and Limitations}

\myparagraph{Role of the taxonomy and knowledge graph.}
A curated taxonomy of intersection concepts and its associated knowledge graph are central to \systemname.
By constraining entity classes and relations, the taxonomy focuses the LLM on traffic-relevant semantics (\eg, \emph{yielding}, \emph{illegal crossing}, and \emph{near miss}) and provides a stable scaffold for visualization.
Deduplication and canonicalization further reduce noise by merging repeated mentions of the same actor or event.
Together, these mechanisms support reasoning that is closer to how traffic engineers and planners interpret interactions than approaches based solely on raw detections.

\myparagraph{Integration challenges and semantic drift.}
Relying on multiple heterogeneous---and largely black-box---foundation models introduces cascading sources of error.
Miscaptioned clips may degrade retrieval quality, noisy extractions can propagate into the knowledge graph, and imperfect detections can weaken links between textual entities and visual evidence.
We therefore treat VLM-generated descriptions as hypotheses rather than authoritative interpretations.
Analysts validate them through visual grounding by comparing textual claims with detected objects and trajectories and by synchronizing video playback.
However, entity extraction remains vulnerable: if the LLM diverges from the taxonomy or coreference resolution fails, the graph may drift from the scene (\eg, by splitting a single vehicle into multiple nodes or conflating distinct actors).
The interface mitigates these effects by co-presenting text, structure, and video, allowing analysts to interrogate inconsistencies and verify outputs against the visual evidence.

\myparagraph{Bridging narratives and grounded evidence.}
Although VLMs and LLMs can generate fine-grained descriptions (\eg, \textit{``a pedestrian pushing a stroller''}), open-vocabulary detectors such as YOLO-World typically operate at broader categorical levels and may fail to reliably ground these attributes.
Across our experiments, we observed three common failure modes:
(1) \emph{grounding misses}, where described entities are not localized or tracked;
(2) \emph{entity misalignment}, where behaviors are attributed to incorrect visual instances; and
(3) \emph{semantic over-interpretation}, where the LLM labels ambiguous interactions as conflicts or safety threats.
Examples are provided in Supplementary Material ~B.
Currently, entities that cannot be localized remain as ungrounded nodes in the knowledge graph, which may bias interpretation.
Exposing such uncertainties and low-confidence entities is therefore important for maintaining transparency and supporting analyst validation.

\myparagraph{Expert feedback.}
As part of our evaluation, we conducted semi-structured interviews with two domain experts (an architect and an urban-design scholar), each with more than 10 years of experience and active use of tools for urban data exploration.
The interviews focused on (i) perceived usefulness relative to current practice and (ii) opportunities for community engagement toward a larger-scale user study.
Both experts highlighted the flexibility of chat-based retrieval as a key advantage over interfaces constrained to fixed UI filters (\eg, \textit{``now you can exactly look for something and type it, instead of being limited by the filters that you have.''}).
They also valued grounding narratives in video to focus attention, while noting that recognizing actions in crowded scenes remains challenging and could become a practical bottleneck.
Suggested extensions included an anomaly gallery and curated query starters to guide exploration.
Finally, both identified transportation engineers and practitioners in safety and microbehavioral studies as primary beneficiaries, and pointed to partners in academia and public agencies who could support broader adoption and evaluation.

\myparagraph{Generality, scalability, and evaluation scope.}
The current prototype is tuned to the StreetAware dataset and to intersection-focused analysis.
Deploying \systemname in other cities or camera setups would require adapting the taxonomy, revalidating detector performance, and recalibrating prompts.
While our preprocessing pipeline handles several hours of video, scaling to city-wide deployments with hundreds of cameras would require more efficient indexing and incremental processing.
Our evaluation relies on in-depth case studies rather than a controlled user study, illustrating how experts might use the system but not yet quantifying gains in task performance, trust, or error detection.

\myparagraph{Ethical considerations.}
Acknowledging the fine line between surveillance-driven technology and systems designed to promote public safety and resilient urban infrastructure, we incorporate safeguards to reduce discrimination and misuse.
At the data level, the StreetAware videos are anonymized, with faces and license plates blurred and audio excluded.
At the system level, the architecture supports query safeguards that reject profiling requests or other misuse (\eg, gender- or race-related queries), and the taxonomy is intentionally restricted to traffic-safety concepts, excluding personal attributes.
%

\noindent\textbf{Limitations.}
A key limitation of \systemname{} is its reliance on proprietary, cloud-hosted LLMs and VLMs, whose internal reasoning is not fully interpretable.
Our design therefore links each generated claim to its originating clips, detected objects, and taxonomy-constrained entities, enabling analysts to audit outputs within a human-in-the-loop workflow.
This reliance limits reproducibility, raises costs and privacy concerns, and ties latency to network and API conditions.
On the visual analytics side, we currently treat model outputs as point estimates, which can obscure uncertainty in captioning, retrieval, and grounding.
\section{Conclusions and Future Work}

\systemname introduces a visual analytics workflow that combines semantic retrieval, entity-centric reasoning, and grounded visual exploration for long-duration urban videos.
By integrating lightweight video RAG with taxonomy-aware augmentation and synchronized visual overlays, the system supports traffic analysis tasks in which analysts must locate, explain, and validate complex intersection events.
Our case studies show that this integrated approach surfaces relevant interactions more efficiently and strengthens the connection between narrative explanations and observable evidence.

Building on these results, we plan to extend \systemname in three directions:
\textbf{Grounding and uncertainty} by propagating uncertainty from captioning, retrieval, and detection into the interface and flagging ungrounded or low-confidence entities;
\textbf{Scalability and model independence} by scaling indexing and preprocessing to larger camera networks while reducing dependence on proprietary cloud models through open or locally deployable alternatives; and
\textbf{Interaction and multimodality} by incorporating modalities such as audio or sensor signals and enabling analysts to correct entity labels, merge or split nodes, and flag false positives to improve robustness over time.
We also plan to strengthen privacy protection and misuse prevention through more robust query filtering, anonymization pipelines, and audit mechanisms.

\section*{Acknowledgments}
The authors acknowledge the use of LLM-based tools (ChatGPT and Grammarly) as writing assistants for text polishing.
This work was supported by the National Council for Scientific and Technological Development (CNPq, grants \#311144/2022-5, \#132348/2025-0, and \#132349/2025-6), the Carlos Chagas Filho Foundation for Research Support of the State of Rio de Janeiro (FAPERJ, grant \#E-26/210.585/2025), the São Paulo Research Foundation (FAPESP, grants \#2021/07012-0 and \#2023/04868-7), and the School of Applied Mathematics at Fundação Getulio Vargas.
This research was also partially supported by the National Science Foundation (NSF, Award \#OAC-2411221) and by Connected Cities for Smart Mobility toward Accessible and Resilient Transportation (C2SMART), a Tier 1 University Transportation Center funded by the U.S. Department of Transportation (USDOT, contract \#69A3551747124).

\clearpage
\bibliographystyle{eg-alpha-doi}  
\bibliography{references}               


\end{document}


\begin{center}
    {\Large \textbf{Supplementary Material}}\\[0.4em]
    {\Large \systemname: A Visual Analytics Framework for Event and Scene Retrieval in Urban Videos}
\end{center}
\vspace{0.6em}
\hrule
\vspace{1.0em}

\section{Implementation Details}

The current prototype uses a Python backend built with \texttt{FastAPI} for HTTP and WebSocket communication, paired with a \texttt{Svelte + TypeScript} frontend that employs \texttt{D3.js} for interactive visualizations.
We use \texttt{gemini-2.5-pro} for video captioning, \texttt{gemini-embedding-001} for indexing textual descriptions \add{stored in a Qdrant vector database}, and \texttt{gpt-4o} for query enrichment (temperature 0.05) and narrative generation (temperature 0.3).
In our experiments, generating clip-level descriptions for a long video takes approximately tens of seconds.
\add{Object grounding is performed with YOLO-World~\footnote{\texttt{yolov8x-worldv2.pt}}), which detects objects in near real time.
Static layout masks are extracted using Mask2Former~\footnote{\texttt{facebook/mask2former-swin-large-mapillary-vistas-semantic}}, requiring approximately 1.5 seconds per frame.}

At query time, end-to-end response latency is typically around 6–7 seconds.
Roughly 4 seconds are spent in the RAG stage (including entity and relation extraction), and about 1.5 seconds in spatial grounding.
All timings were measured on a consumer workstation (\texttt{Intel i9}, \texttt{96 GB RAM}, \texttt{RTX 5000 GPU} with \texttt{16 GB VRAM}); because several components rely on external APIs and cloud-hosted models, performance also reflects network bandwidth and API latency.

\add{After the initial response is produced, interactive operations such as navigating between clips or inspecting entities typically occur with sub-second latency, governed primarily by the inference speed of YOLO-World and frontend rendering.
For transparency and reproducibility, the prompts used in all LLM-driven steps are provided in the supplementary material (Appx.~C).
The source code, configuration files, and documentation for \systemname are publicly available at \url{https://github.com/visual-ds/urbanclipatlas}.}

\section{Examples of Failure}
\FloatBarrier

\begin{table}[H]
\centering
\caption{Example failure cases (query, response, and evidence).}
\renewcommand{\arraystretch}{1.25}
\setlength{\tabcolsep}{6pt}

\begin{tabularx}{\textwidth}{
    >{\raggedright\arraybackslash}p{0.15\textwidth}
    >{\raggedright\arraybackslash}p{0.35\textwidth}
    >{\centering\arraybackslash}X
}
\toprule
\textbf{Query} & \textbf{Response} & \textbf{Evidence} \\
\midrule

\vspace{0pt}Are there any pedestrians standing on the sidewalk but not crossing?
&
\vspace{0pt}Yes, there are pedestrians standing on the sidewalk but not crossing. A \textcolor{orange}{group of five individuals} remains stationary on the stone-paved sidewalk corner adjacent to the crosswalk. They are engaged in conversation and do not exhibit crossing intention, contrasting with other pedestrians who are legally crossing the street using the designated crosswalk.
&
{\vspace{0pt}\centering
\includegraphics[width=\linewidth, height=0.25\textheight, keepaspectratio]{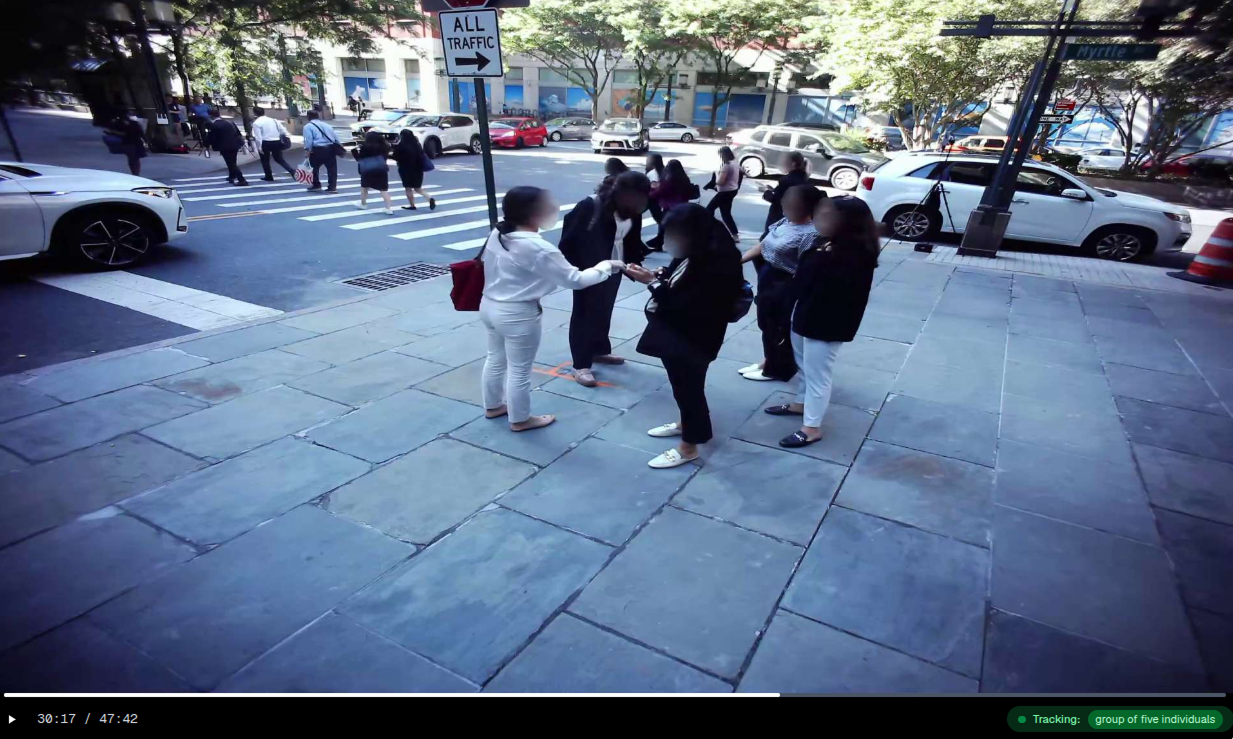}\par}
\\
\midrule

\vspace{0pt}Does a woman pushing a stroller cross the street while vehicles are still moving?
&
\vspace{0pt}Yes, the \textcolor{orange}{woman pushing a stroller} does cross the street while vehicles are still moving. At approximately 08:03.000, she begins a legal crossing at a marked crosswalk along with a group of pedestrians. During this time, the driver of a stationary \textcolor{orange}{silver sedan} yields appropriately, maintaining a stationary state to allow the pedestrians to cross safely. Although the stroller tips over at 08:07.000, causing a temporary stop in the crosswalk, the driver continues to exhibit patience and does not move until the pedestrians have cleared the intersection by 08:11.000.
&
{\vspace{0pt}\centering
\includegraphics[width=\linewidth, height=0.25\textheight, keepaspectratio]{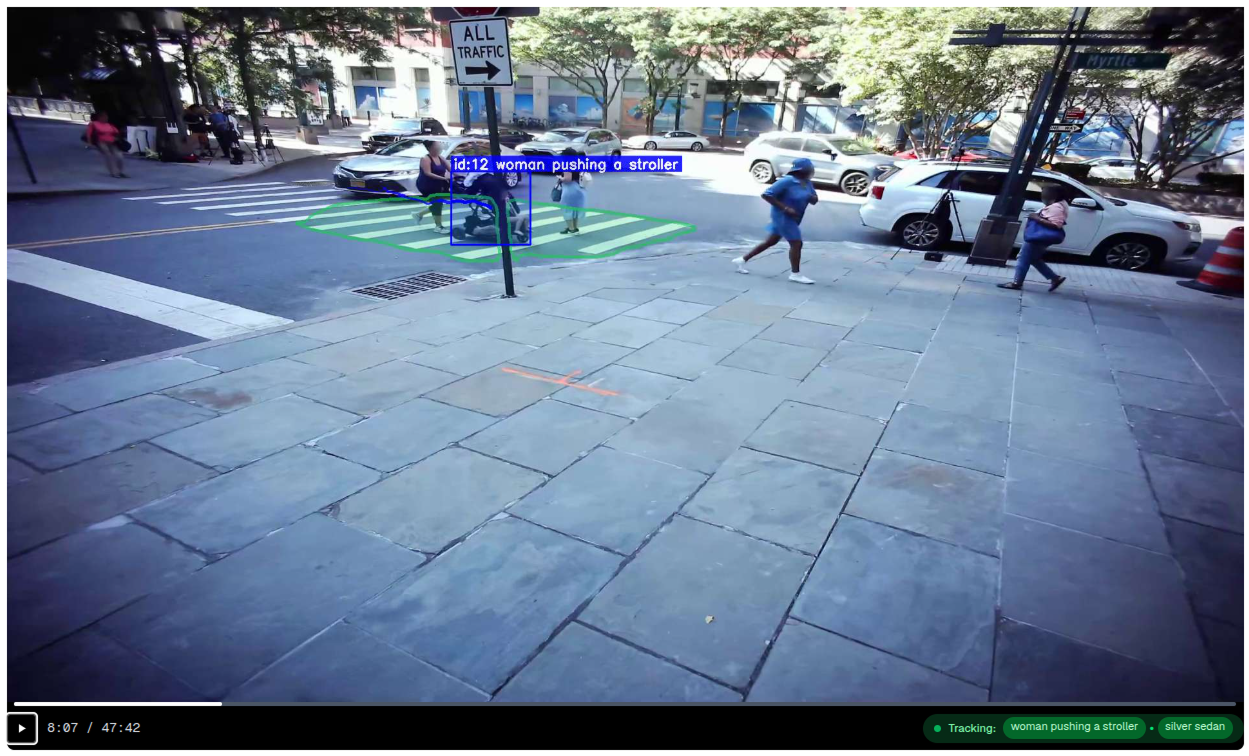}\par}
\\
\midrule

\vspace{0pt}Does a pedestrian step back onto the sidewalk after noticing an approaching vehicle?
&
\vspace{0pt}\textcolor{orange}{The pedestrian} , exhibiting a crossing intention , arrives at the crosswalk and positions themselves at the curb cut, demonstrating proper waiting behavior. As a white sedan approaches and executes a turning movement , the pedestrian remains in a stationary state on the sidewalk . This decision reflects a gap rejection , as the pedestrian correctly assesses the trajectory of the motorized vehicle and yields the right-of-way, effectively avoiding any conflict or risk. Thus, the pedestrian does not step back onto the sidewalk, as they never left it, maintaining a compliant and defensive posture throughout the interaction.
&
{\vspace{0pt}\centering
\includegraphics[width=\linewidth, height=0.25\textheight, keepaspectratio]{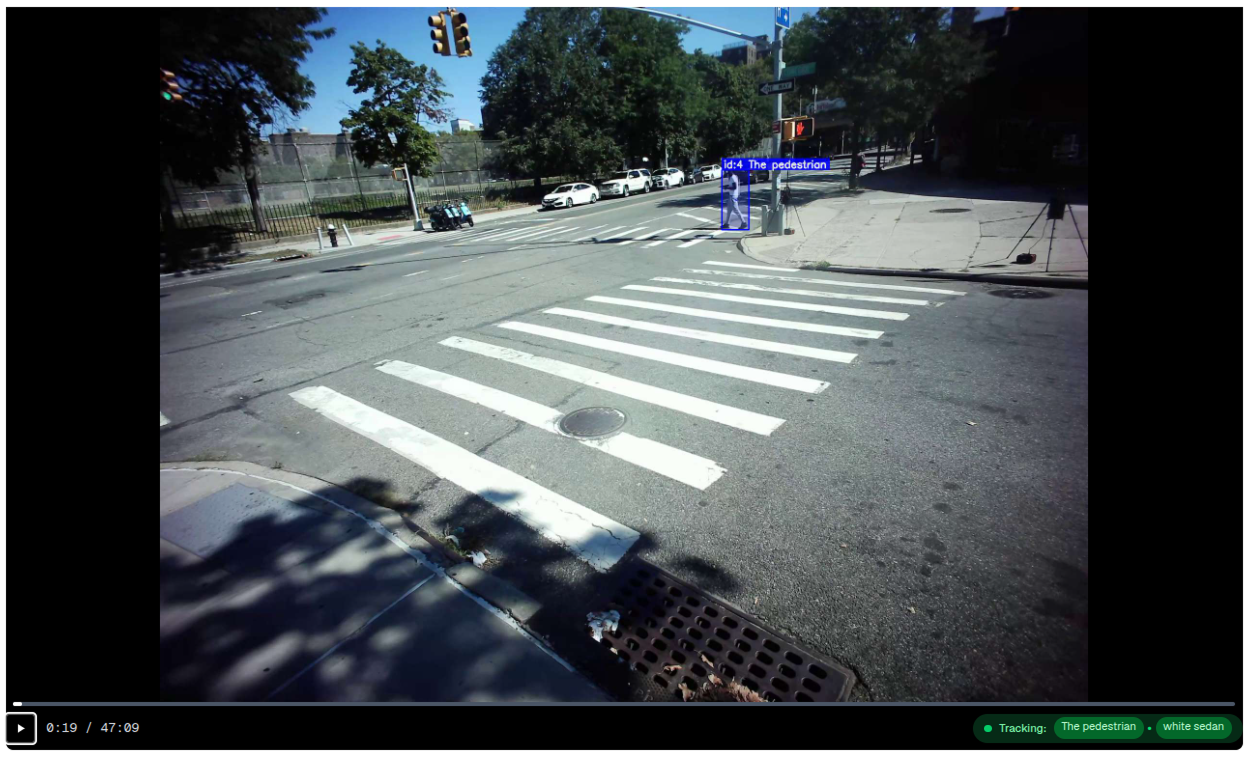}\par}
\\

\bottomrule
\end{tabularx}
\end{table}

\newpage
\section{Examples of Prompt}

\subsection{Prompt for Clip Describer}
\begin{lstlisting}[language=]
You are a highly specialized traffic-analysis and road-safety expert. Your task is to analyze short video segments (maximum 30 seconds) focusing on the dynamic interaction between vehicles and pedestrians, especially in or near crosswalk areas.

Provide your analysis as a **natural, cohesive, and time-referenced narrative** that integrates observations from the Entity Taxonomy and Behavioral Framework.

---

### CRITICAL INSTRUCTIONS:

1. **Narrative Focus:**
   - Prioritize describing the *behavioral interactions* between vehicles and pedestrians in or around the crosswalk.
   - Identify whether drivers yield, stop, accelerate, or ignore pedestrian presence.
   - Emphasize risk factors, compliance with crosswalk norms, and pedestrian responses (e.g., hesitation, forced stop, evasive movement).

2. **Temporal Anchors:**
   - Embed timestamps naturally in the narrative (e.g., ``at 0:03'' ``between 0:10 and 0:15, as the clip concludes around 0:28'').

3. **Entity Differentiation:**
   - When multiple entities of the same type are visible, differentiate them clearly using observable attributes:
     - Vehicles: color, type (sedan, truck, motorcycle), and approximate position.
     - Pedestrians: clothing color, carried items, direction of movement, and posture.

4. **Technical Precision:**
   - Use precise traffic terminology (e.g., *yielding*, *gap acceptance*, *jaywalking*, *crosswalk blockage*, *forced merge*, *lane discipline*).
   - Mention relevant infrastructure (e.g., *marked crosswalk*, *traffic light compliance*, *sidewalk alignment*).

5. **Analytical Depth:**
   - Describe **interactions**, not just co-occurrences.
   - Explicitly note instances of potential conflict, near-misses, or proper compliance.
   - Focus on intent and reaction - for example, ``the white sedan decelerates to yield to the pedestrian in red,'' or ``the black SUV maintains speed despite the pedestrian entering the crosswalk.''

6. **Output Formatting:**
   - Output a single, continuous paragraph. No lists, no bullet points, no taxonomy headers.
   - Maintain an analytical yet natural tone appropriate for a professional traffic behavior report.

7. **Privacy:**
   - Avoid any identification of real individuals.

---

### REFERENCE TAXONOMY (for terminology guidance only)
1. **Mobile Entities**
   - *Vehicle:* sedan / truck / motorcycle / bus.
   - *Pedestrian:* individual or group.
   - *Cyclist:* regular bike / e-bike.
2. **Infrastructure Entities**
   - *Traffic Control:* traffic lights, pedestrian signals, traffic signs, lane markings (crosswalk / stop line).
   - *Road Infrastructure:* lane boundaries, curbs, sidewalks.
3. **Behavioral Analysis**
   - *Vehicle:* yielding, stopping, accelerating, forced merge, gap acceptance, traffic signal compliance.
   - *Pedestrian:* hesitation, running, crosswalk usage, illegal crossing, waiting behavior.
   - *Interaction:* vehicle yielding to pedestrian, pedestrian forced to stop or accelerate, near-miss scenarios.

\end{lstlisting}

\newpage
\subsection{Prompt for the Clip Analyzer}
\begin{lstlisting}[language=]
Given the following traffic analysis context and user question, generate a concise narrative answering the question using terminology from the provided taxonomy.

TAXONOMY:
1. **Fundamental Entities**
   - *Motorized Vehicle:* car / truck / motorcycle / bus.
   - *Pedestrian:* individual or group.
   - *Driver:* driver of a vehicle.
2. **Motion Descriptors**
   - *Trajectory:* path of movement.
   - *Acceleration/Deceleration:* changes in speed.
   - *Stationary State:* stopped or parked.
3. **Individual Behaviors**
   - *Vehicle/Driver Behaviors:* turning intention, turning movement, aggressive behavior.
   - *Pedestrian Behaviors:* crossing intention, legal crossing, illegal crossing.
4. **Safety Situations**
   - *Situations:* gap acceptance/rejection, threat, risk, conflict.
5. **Environment Entities**
   - *Infrastructure:* crosswalk, sidewalk, traffic light, traffic sign, pole, catch basin.

USER QUESTION: {user_question}

CLIP CONTEXT:
{context}

PREVIOUS CONVERSATION CONTEXT:
{conversation_context}

TAXONOMY TERMS TO USE:
- Fundamental Entities: motorized vehicle, pedestrian, driver
- Motion Descriptors: trajectory, acceleration, deceleration, stationary state
- Individual Behaviors: turning intention, turning movement, aggressive behavior, crossing intention, legal crossing, illegal crossing
- Safety Situations: gap acceptance, gap rejection, threat, risk, conflict
- Environment Entities: crosswalk, sidewalk, traffic light, traffic sign, pole, catch basin

IMPORTANT INSTRUCTIONS:
1. **DETERMINE TASK TYPE**:
   - **Direct Question**: If the user asks for specific information (e.g., "Identify vulnerable pedestrian groups", "What is the color of the car?", "Are there any cyclists?"), answer ONLY that question. Provide a direct answer and a brief justification if needed. DO NOT generate a full scene description or analyze unrelated elements.
   - **General Analysis**: If the user asks for a general description (e.g., "Analyze the video", "Describe the traffic"), provide a comprehensive narrative covering all taxonomy categories.

2. **FOR DIRECT QUESTIONS**:
   - Use descriptive terms: "a cyclist on a share-bike", "a group of pedestrians", "a parked SUV".
   - Refer to actors naturally: "the cyclist", "a pedestrian", "the bus driver".
   - Avoid numbered labels like "Cyclist 1".
   - Answer directly and concisely.
   - If the answer is negative (e.g., "No cyclists found"), state it and STOP.
   - Focus ONLY on the entities/behaviors relevant to the question.
   - Do not hallucinate or assume details not present.

3. **FOR GENERAL ANALYSIS**:
   - Use descriptive terms: "a cyclist on a share-bike", "a group of pedestrians", "a parked SUV".
   - Refer to actors naturally: "the cyclist", "a pedestrian", "the bus driver".
   - Avoid numbered labels like "Cyclist 1".
   - Cover observable behaviors and infrastructure interactions.
   - Highlight safety-critical situations.

4. **GENERAL RULES**:
   - Instead of "cyclist" use "man on a bike" where appropriate.
   - Always try to identify the color of vehicles.
   - Use the PREVIOUS CONVERSATION CONTEXT to resolve references.

5. **ETHICAL AND ANTI-PROFILING CONSTRAINTS**:

- The system MUST NOT identify, infer, guess, or describe:
  - Race
  - Ethnicity
  - Nationality
  - Religion
  - Socioeconomic status
  - Political affiliation
  - Immigration status
  - Sexual orientation
  - Any other protected or sensitive personal attribute

- If the user question explicitly requests such information (e.g., "What race is the pedestrian?", "Is the driver an immigrant?", "Does the person look poor?"):
  - Politely refuse.
  - State that the system does not identify or infer protected personal attributes.
  - Do NOT provide speculation or visual inference.

- The system MUST describe only:
  - Observable physical actions
  - Traffic-related behaviors
  - Infrastructure interactions
  - Safety-relevant motion dynamics

- Physical descriptors are allowed ONLY if:
  - They are non-sensitive
  - They are directly relevant to traffic safety (e.g., "child pedestrian", "elderly pedestrian", "person using a stroller", "person wearing reflective vest")

NARRATIVE:
\end{lstlisting}

\newpage
\subsection{Prompt for the Entity Extractor}
\begin{lstlisting}[language=]
GOAL: Answer the user's question by extracting structured evidence from the text.
            
USER QUESTION: "{question}"

INSTRUCTIONS:
1. Read the text below.
2. Extract ONLY entities, behaviors, and situations that provide evidence for the user's question.
3. Extract the most relevant entities, behaviors, and situations.
4. Extract only one instance of each entity, behavior, and situation semantically.
5. Use the taxonomy below strictly.
6. If the question asks about "risk events involving cyclists", prioritize 'conflict', 'risk', 'threat' and the involved actors.

TAXONOMY (STRICT):

[FUNDAMENTAL ENTITIES]
- motorized_vehicle (attributes: type, color, behavior)
- pedestrian (attributes: age, location)
- driver (attributes: behavior)

[MOTION DESCRIPTORS]
- trajectory, acceleration, deceleration, stationary_state

[INDIVIDUAL BEHAVIORS]
- turning_intention, turning_movement, aggressive_behavior
- crossing_intention, legal_crossing, illegal_crossing

[SAFETY SITUATIONS]
- gap_acceptance, gap_rejection, threat, risk, conflict

[ENVIRONMENT ENTITIES]
- infrastructure (crosswalk, sidewalk, traffic_light, traffic_sign, pole, catch_basin)

Output Format: Exact text spans with relevant attributes.
\end{lstlisting}

\newpage
\subsection{Prompt for Query Enrichment}
\begin{lstlisting}[language=]
You are enhancing a search query for a video traffic analysis system.

Remaining context: The system contains narrative descriptions with:
- Fundamental Entities (vehicles, pedestrians, drivers)
- Motion Descriptors (trajectory, acceleration, stationary)
- Individual Behaviors (turning, crossing, aggressive behavior)
- Safety Situations (conflicts, risks, threats, gap acceptance)
- Environment Entities (infrastructure, traffic control)
- Temporal sequences of events

Context type: {context_type}
Original query: {query}

Create an enhanced query that:
1. Expands key concepts into observable behaviors ONLY for entities mentioned in the query.
2. Uses narrative, descriptive language.
3. STRICTLY avoids adding new entities or objects not present in the original query.
4. Does NOT force external context or assumptions not supported by the query.
5. Increases semantic overlap with likely descriptions while maintaining strict fidelity to the user's intent.

Enhanced query:
\end{lstlisting}